\documentclass[nofootinbib,reprint,amsmath,amssymb,aps]{revtex4-1}
\usepackage{graphicx}
\usepackage{dcolumn}
\usepackage{xcolor}
\usepackage{bm}
\usepackage{hyperref}
\newcommand{\qm}[1]{``#1''}

\newcommand\RS{{\mathrm{R_\star}}}

\begin{document}

\preprint{APS/123-QED}

\title[Stable attractors in PR effect]{Stable attractors in the three-dimensional\\ general relativistic Poynting-Robertson effect}

\author{Vittorio De Falco$^{1}$},\email{vittorio.defalco@physics.cz}
\author{Pavel Bakala$^{1,2}$}\email{pavel.bakala@physics.cz}\vspace{0.5cm}

\affiliation{$^1$ Research Centre for Computational Physics and Data Processing, Faculty of Philosophy \& Science, Silesian University in Opava, Bezru\v{c}ovo n\'am.~13, CZ-746\,01 Opava, Czech Republic \\
$^2$M. R. \v{S}tef\'anik Observatory and Planetarium, Sl\'adkovi\v{c}ova 41, 920 01 Hlohovec, Slovak Republic}

\date{\today}

\begin{abstract}
We prove the stability of the critical hypersurfaces associated with the three-dimensional general relativistic Poynting-Robertson effect. The equatorial ring configures to be as a stable attractor and the whole critical hypersurface as a basin of attraction for this dynamical system. We introduce a new, simpler (in terms of calculations), and more physical approach within the Lyapunov theory. We propose three different Lyapunov functions, each one carrying important information and very useful for understanding such phenomenon under different aspects.
\end{abstract}

\maketitle
\section{Introduction}
\label{sec:intro}
In the study of radiation processes in high-energy astrophysics around compact objects, like neutron stars (NSs) and black holes (BHs), it is of utmost importance to accurately describe the motion of the surrounding matter to then test the model in strong field regimes. In particular, when we deal with relatively small-sized test particles, like dust grains, plasma or gas elements invested by an electromagnetic radiation field, their motion can be considerably altered by the general relativistic Poynting-Robertson (PR) effect \cite{Poynting1903,Robertson1937}. The radiation field exerts not only a force outward the compact object contrasting the gravitational pull, but also a radiation drag force opposite to the test particle orbital velocity. The PR effect configures thus as a pure relativistic dissipative effect, which efficiently removes energy and angular momentum from the affected test particle. 

The general relativistic treatments, from the two dimensional (2D) \cite{Bini2009,Bini2011} until the three dimensional (3D) formulations \cite{DeFalco20183D,Bakala2019,Wielgus2019}, show all the existence of a critical hypersurface, region where gravitational and radiation forces balance. From selected test particle orbits, it graphically results that the critical hypersurfaces are stable, namely once the test particle reaches such region and moves on that, it will remain there forever. The test particle can either spiral down towards the equatorial ring (latitudinal drift) or move in suspended orbits (see Refs. \cite{DeFalco20183D,Bakala2019}, for details). This implication must be formally proved, because it contains fundamental information not only on the PR effect, but also on the system under study. 

We propose a new method to prove the stability of the critical hypersurfaces within the \emph{Lyapunov theory}. Such approach carries important information on the physical system under study and it substantially reduces the calculations with respect to a previous method employed in the literature \cite{Bini2011}. The idea of a Lyapunov function has been proposed in 1956 \cite{Lyapunov1992}, and since then it has been exploited in several and disparate contexts in physics and in mathematics \cite{Lasalle1961,OGUZTORELI1981,Robert2004,Hahn2019}. There are also a wealth of applications in astrophysics and cosmology, such as: in accretion disc theory to control the large- and small-scale in/stability of such continuous-like structures (see, e.g., Refs. \cite{Mineshige1993,Hawley1999,Ross2017,Kadowaki2018});  
in celestial mechanics to study the motion of bodies under the influence of a gravitational (or other kinds of) forces (see, e.g., Refs. \cite{Singh2010,Blaga2016,LIU2019}); in cosmology to understand the stability of the models, to analyse dark energy's origin and implications, and to investigate modified gravity scenarios (see e.g., Refs \cite{Charters2001,Bahamonde2018}).

The paper is structured as follows: in Sec. \ref{sec:PRmodel} we introduce the formal aspects of the 3D general relativistic PR effect model, including a detailed description of the critical hypersurfaces' derivation and proprieties. In Sec. \ref{sec:stability} we review what has been done in the literature so far, underlying the limiting aspects, and then we present the power and advantages of our new approach. Finally, in Sec. \ref{sec:end} we draw our conclusions. 

\section{General relativistic 3D PR effect model}
\label{sec:PRmodel}
We consider a central compact object, whose outside spacetime is described by the Kerr metric with signature $(-,+,+,+)$. In geometrical units ($c = G = 1$), the line element of the Kerr spacetime, $ds^2=g_{\alpha\beta}dx^\alpha dx^\beta$, in Boyer-Lindquist coordinates, parameterized by mass $M$ and spin $a$, reads as
\begin{equation}\label{kerr_metric}
\begin{aligned}
 \mathrm{d}s^2 &= -\left(1-\frac{2Mr}{\Sigma}\right) \,\mathrm{d}t^2 
  - \frac{4Mra}{\Sigma} \sin^2\theta\,\mathrm{d}t\, \mathrm{d}\varphi \\
  &+ \frac{\Sigma}{\Delta}\, \mathrm{d}r^2 
  + \Sigma \,\mathrm{d}\theta^2
  + \rho\sin^2\theta\, \mathrm{d}\varphi^2, 
\end{aligned}
\end{equation}
where $\Sigma \equiv r^{2} + a^{2}\cos^{2}\theta$, $\Delta \equiv r^{2} - 2Mr + a^{2}$, and $\rho  \equiv r^2+a^2+2Ma^2r\sin^2\theta/\Sigma$. The determinant of the Kerr metric is $g=-\Sigma^2\sin^{2}\theta$. We introduce the zero angular momentum observers (ZAMOs), whose adapted orthonormal frame is given by
\begin{equation} \label{eq:zamoframes}
\begin{aligned}
&\boldsymbol{e_{\hat t}}\equiv\boldsymbol{n}=\frac{1}{N}(\boldsymbol{\partial_t}-N^{\varphi}\boldsymbol{\partial_\varphi}),\quad
\boldsymbol{e_{\hat r}}=\frac1{\sqrt{g_{rr}}}\boldsymbol{\partial_r},\\
&\boldsymbol{e_{\hat \theta}}=\frac1{\sqrt{g_{\theta \theta }}}\boldsymbol{\partial_\theta},\quad
\boldsymbol{e_{\hat \varphi}}=\frac1{\sqrt{g_{\varphi \varphi }}}\boldsymbol{\partial_\varphi},
\end{aligned}
\end{equation}
where where $\left\{\boldsymbol{\partial_t},\ \boldsymbol{\partial_r},\ \boldsymbol{\partial_\theta},\ \boldsymbol{\partial_\varphi}\right\}$ is the orthonormal frame adapted to the static observer at infinity, $N=(-g^{tt})^{-1/2}$ is the time lapse function and $N^{\varphi}=g_{t\varphi}/g_{\varphi\varphi}$ the spatial shift vector field. All the vector and tensor indices (e.g., $v^\alpha, T^{\alpha\beta}$) associated to the ZAMO frame will be labeled by a hat (e.g., $v^{\hat\alpha}, T^{\hat{\alpha}\hat{\beta}}$), instead all the scalar quantities measured in the ZAMO frame (e.g., $f$) will be followed by $(n)$ (e.g., $f(n)$). In the kinematical decomposition of the ZAMO congruence, we have that the nonzero ZAMO kinematical quantities are acceleration $\boldsymbol{a}(n)=\nabla_{\boldsymbol{n}} \boldsymbol{n}$, expansion tensor along the $\hat{\varphi}$-direction $\boldsymbol{\theta_{\hat\varphi}}(n)$, and the relative Lie curvature vector $\boldsymbol{k_{(\rm Lie)}}(n)$ (see Table 1 in \cite{DeFalco20183D}, for their explicit expression).
 
The radiation field is constituted by a coherent flux of photons traveling along null geodesics in the Kerr geometry. The related stress-energy tensor is \cite{DeFalco20183D,Bakala2019}
\begin{equation}\label{STE}
T^{\mu\nu}=\mathcal{I}^2 k^\mu k^\nu\,,\qquad k^\mu k_\mu=0,\qquad k^\mu \nabla_\mu k^\nu=0,
\end{equation}
where $\mathcal{I}$ is a parameter linked to the radiation field intensity and $\boldsymbol{k}$ is the photon four-momentum field. Splitting $\boldsymbol{k}$ with respect to the ZAMO frame, we obtain
\begin{eqnarray}
&&\boldsymbol{k}=E(n)[\boldsymbol{n}+\boldsymbol{\hat{\nu}}(k,n)], \label{photon1}\\
&&\boldsymbol{\hat{\nu}}(k,n)=\sin\zeta\sin\beta\ \boldsymbol{e_{\hat r}}+\cos\zeta\ \boldsymbol{e_{\hat\theta}}+\sin\zeta \cos\beta\ \boldsymbol{e_{\hat\varphi}}, \label{photon2}
\end{eqnarray}
where $\boldsymbol{\hat{\nu}}(k,n)$ is the photon spatial unit relative velocity with respect to the ZAMOs, $\beta$ and $\zeta$ are the two angles measured in the ZAMO frame in the azimuthal and polar direction, respectively. In addition, $E(n)$ is the photon energy measured in the ZAMO frame, which is \cite{DeFalco20183D,Bakala2019}
\begin{equation}
E(n)=\frac{E_p}{N}(1+bN^\varphi),
\end{equation}
where $E_p=-k_t$ is the conserved photon energy along its trajectory. The radiation field is governed by the two photon impact parameters $(b,q)$, associated respectively with the two emission angles $(\beta,\zeta)$. 

The photons of the radiation field are emitted from a spherical surface having a radius $\RS$ centered at the origin of the Boyer-Lindquist coordinates, and rigidly rotating with angular velocity $\Omega_{\mathrm{\star}}$. The photon impact parameters have the following expressions \cite{Bakala2019}
\begin{eqnarray} 
&&b=-\left[\frac{\mathrm{g_{t\varphi}}+\mathrm{g_{\varphi\varphi}}\Omega_{\star} }{\mathrm{g_{tt}}+\mathrm{g_{t\varphi}} \Omega_{\star}}\right]_{r=\RS},\label{kerr_impact_parameter}\\  
&&q=\left[b^{2}\cot^{2} \theta-a^{2} \cos^{2}\theta \right]_{r=\RS}. \label{q_r}
\end{eqnarray}
The related photon angles in the ZAMO frame are \cite{Bakala2019}
\begin{equation} \label{ANG1}
\cos\beta=\frac{b N}{\sqrt{g_{\varphi\varphi}}(1+b N^\varphi)}, \qquad \zeta=\pi/2. 
\end{equation}
The parameter $\mathcal{I}$ is given by \cite{Bakala2019}
\begin{equation}\label{INT_PAR}
\mathcal{I}^2=\frac{\mathcal{I}_0^2}{\sqrt{\left( r^{2} + a^{2}-ab \right)^{2}- \Delta \left[ q + \left( b - a \right) ^{2} \right]}},
\end{equation}
where $\mathcal{I}_0$ is $\mathcal{I}$ evaluated at the emitting surface.

A test particle moves with a timelike four-velocity $\boldsymbol{U}$ and a spatial three-velocity with respect to the ZAMOs, $\boldsymbol{\nu}(U,n)$, which both read as \cite{DeFalco20183D}
\begin{eqnarray} 
&&\boldsymbol{U}=\gamma(U,n)[\boldsymbol{n}+\boldsymbol{\nu}(U,n)], \label{testp}\\
&&\boldsymbol{\nu}=\nu(\sin\psi\sin\alpha\boldsymbol{e_{\hat r}}+\cos\psi\boldsymbol{e_{\hat\theta}}+\sin\psi \cos\alpha \boldsymbol{e_{\hat\varphi}}),
\end{eqnarray}
where $\gamma(U,n)\equiv\gamma=1/\sqrt{1-||\boldsymbol{\nu}(U,n)||^2}$ is the Lorentz factor, $\nu=||\boldsymbol{\nu}(U,n)||$, $\gamma(U,n) =\gamma$. We have that $\nu$ represents the magnitude of the test particle spatial velocity $\boldsymbol{\nu}(U,n)$, $\alpha$ is the azimuthal angle of the vector $\boldsymbol{\nu}(U,n)$ measured clockwise from the positive $\hat\varphi$ direction in the $\hat{r}-\hat{\varphi}$ tangent plane in the ZAMO frame, and $\psi$ is the polar angle of the vector $\boldsymbol{\nu}(U,n)$ measured from the axis orthogonal to the $\hat{r}-\hat{\varphi}$ tangent plane in the ZAMO frame. 

We assume that the radiation test particle interaction occurs through Thomson scattering, characterized by a constant momentum-transfer cross section $\sigma$, independent from direction and frequency of the radiation field. The radiation force is given by \cite{DeFalco20183D,Bakala2019}
\begin{equation} \label{radforce}
{\mathcal F}_{\rm (rad)}(U)^{\hat \alpha}=\sigma \, [\mathcal{I} E(U)]^2\, \hat{\mathcal V}(k,U)^{\hat \alpha}\,.
\end{equation}
where the term $\tilde{\sigma}[\mathcal{I} E(U)]^2$ reads as \cite{DeFalco20183D,Bakala2019} 
\begin{equation} \label{eq: sigma_tilde}
\tilde{\sigma}[\mathcal{I} E(U)]^2=\frac{ A\,\gamma^2(1+b N^\varphi)^2[1-\nu\sin\psi\cos(\alpha-\beta)]^2}{N^2\sqrt{\left( r^{2} + a^{2}-ab \right)^{2}- \Delta \left[ q + \left( b - a \right) ^{2} \right]}}.
\end{equation}
$E(U)$ is the photon energy absorbed by the test particle, which can be related to $E(n)$ through \cite{DeFalco20183D,Bakala2019}
\begin{equation} \label{enepart}
E(U)=\gamma E(n)[1-\nu\sin\psi\cos(\alpha-\beta)].
\end{equation}

The term $A=\tilde{\sigma}[\mathcal{I}_0 E_p]^2$ is the luminosity parameter, which can be equivalently written $A/M=L/L_{\rm EDD}\in[0,1]$ with $L$ the emitted luminosity at infinity and $L_{\rm EDD}$ the Eddington luminosity. The terms $\hat{\mathcal V}(k,U)^\alpha$ are the radiation field components, which are \cite{DeFalco20183D,Bakala2019}
\begin{eqnarray}
&&\hat{\mathcal{V}}^{\hat r}=\frac{\sin\beta}{\gamma [1-\nu\sin\psi\cos(\alpha-\beta)]}-\gamma\nu\sin\psi\sin\alpha,\label{rad1}\\
&&\hat{\mathcal{V}}^{\hat \theta}=-\gamma\nu\cos\psi \label{rad2},\\
&&\hat{\mathcal{V}}^{\hat\varphi}=\frac{\cos\beta}{\gamma [1-\nu\sin\psi\cos(\alpha-\beta)]}-\gamma\nu\sin\psi\cos\alpha,\label{rad3}\\
&&\hat{\mathcal{V}}^{\hat t}=\gamma\nu\left[\frac{\sin\psi\cos(\alpha-\beta)-\nu}{1-\nu\sin\psi\cos(\alpha-\beta)}\right].\label{rad4}
\end{eqnarray}

Collecting all the information together, it is possible to derive the resulting equations of motion for a test particle moving in a 3D space, which are \cite{DeFalco20183D,Bakala2019}
\begin{eqnarray}
&&\frac{d\nu}{d\tau}= -\frac{1}{\gamma}\left\{ \sin\alpha \sin\psi\left[a(n)^{\hat r}\right.\right.\label{EoM1}\\
&&\left.\left.\ \quad +2\nu\cos \alpha\sin\psi\, \theta(n)^{\hat r}{}_{\hat \varphi} \right]+\cos\psi\left[a(n)^{\hat \theta}\right. \right.\nonumber\\
&&\left.\left.\ \quad+2\nu\cos\alpha\sin\psi\, \theta(n)^{\hat \theta}{}_{\hat \varphi}\right]\right\}+\frac{\tilde{\sigma}[\Phi E(U)]^2}{\gamma^3\nu}\hat{\mathcal{V}}^{\hat t},\nonumber
\end{eqnarray}
\begin{eqnarray}
&&\frac{d\psi}{d\tau}= \frac{\gamma}{\nu} \left\{\sin\psi\left[a(n)^{\hat \theta}+k_{\rm (Lie)}(n)^{\hat \theta}\,\nu^2 \cos^2\alpha\right.\right.\label{EoM2}\\
&&\left.\left.\ \quad+2\nu\cos \alpha\sin^2\psi\ \theta(n)^{\hat \theta}{}_{\hat \varphi}\right]-\sin \alpha\cos\psi \left[a(n)^{\hat r}\right.\right.\nonumber\\
&&\left.\left.\ \quad+k_{\rm (Lie)}(n)^{\hat r}\,\nu^2+2\nu\cos \alpha\sin\psi\, \theta(n)^{\hat r}{}_{\hat \varphi}\right]\right\}\nonumber\\
&&\ \quad+\frac{\tilde{\sigma}[\Phi E(U)]^2}{\gamma\nu^2\sin\psi}\left[\hat{\mathcal{V}}^{\hat t}\cos\psi-\hat{\mathcal{V}}^{\hat \theta}\nu\right],\nonumber\\
&&\frac{d\alpha}{d\tau}=-\frac{\gamma\cos\alpha}{\nu\sin\psi}\left[a(n)^{\hat r}+2\theta(n)^{\hat r}{}_{\hat \varphi}\ \nu\cos\alpha\sin\psi\right.\label{EoM3}\\
&&\left.\ \quad+k_{\rm (Lie)}(n)^{\hat r}\,\nu^2+k_{\rm (Lie)}(n)^{\hat \theta}\,\nu^2\cos^2\psi \sin\alpha\right]\nonumber\\
&&\ \quad+\frac{\tilde{\sigma}[\Phi E(U)]^2\cos\alpha}{\gamma\nu\sin\psi}\left[\hat{\mathcal{V}}^{\hat r}-\hat{\mathcal{V}}^{\hat \varphi}\tan\alpha\right],\nonumber\\
&&U^r\equiv\frac{dr}{d\tau}=\frac{\gamma\nu\sin\alpha\sin\psi}{\sqrt{g_{rr}}}, \label{EoM4}\\
&&U^\theta\equiv\frac{d\theta}{d\tau}=\frac{\gamma\nu\cos\psi}{\sqrt{g_{\theta\theta}}} \label{EoM5},\\
&&U^\varphi\equiv\frac{d\varphi}{d\tau}=\frac{\gamma\nu\cos\alpha\sin\psi}{\sqrt{g_{\varphi\varphi}}}-\frac{\gamma N^\varphi}{N},\label{EoM6}\\
&&U^t\equiv \frac{dt}{d\tau}=\frac{\gamma}{N},\label{time}
\end{eqnarray}
where $\tau$ is the affine (proper time) parameter along $\bold{U}$. 

\subsection{Critical hypersurfaces}
\label{sec:critc_rad}
The dynamical system defined by Eqs. (\ref{EoM1})--(\ref{EoM6}) exhibits an axially symmetric hypersurface outside around the compact object, where there exists a balance among gravitational and radiation forces. We impose that on such region the test particle must move in purely circular orbits with constant velocity ($\nu=\mbox{const}$) with respect to the ZAMO frame ($\alpha=0,\pi$), and the polar axis orthogonal to the critical hypersurface ($\psi=\pm\pi/2$). These requirements entail $d\nu/d\tau=d\alpha/d\tau=0$, from which we determine the following conditions \cite{DeFalco20183D,Bakala2019}
\begin{eqnarray}
&&\nu=\cos\beta, \label{eq:crit_hyper1} \\
&&a(n)^{\hat r}+2\theta(n)^{\hat r}{}_{\hat\varphi}\nu+k_{\rm (Lie)}(n)^{\rm \hat r}\nu^2\label{eq:crit_hyper2}\\
&&=\frac{A(1+bN^\varphi)^2\sin^3\beta}{N^2\gamma\sqrt{\left( r^{2} + a^{2}-ab \right)^{2}- \Delta \left[ q + \left( b - a \right) ^{2} \right]}},\notag 
\end{eqnarray}
where Eq. (\ref{eq:crit_hyper1}) means that the test particle moves on the critical hypersurface with constant velocity equal to the azimuthal photon velocity; whereas Eq. (\ref{eq:crit_hyper2}) determine the critical radius $r_{\rm crit}$ as a function of the polar angle $\theta$ through an implicit equation, once the radiation field proprieties are assigned (i.e., the radius $R_\star$ and the angular velocity $\Omega_\star$ of the emitting surface together with the luminosity parameter $A$, see Ref. \cite{Bakala2019}). It is important to note that Eq. (\ref{eq:crit_hyper2}) might admit three different solutions with precise locations: one inside the emitting surface (considered to be non physical), one close to the emitting surface (the solution we actually study and plot), and another one very far from the emitting surface (that we do not take into account) \cite{Bini2011,Bakala2019}.

In general we have  $d\psi/d\tau\neq0$, because the $\psi$ angle change during the test particle motion on the critical hypersurface, having what we termed \emph{latitudinal drift}. This effect, occurring for the interplay of gravitational and radiation actions in the polar direction, brings definitively the test particle on the equatorial plane \cite{DeFalco20183D,Bakala2019}. For $\psi=\theta=\pi/2$, we have $d\psi/d\tau=0$, corresponding to the equatorial ring. However, we can have $d\psi/d\tau=0$, also for $\theta=\bar{\theta}\neq\pi/2$, having what we termed as \emph{suspended orbits}. The condition for this last configuration in the case of $b\neq0$ can be expressed as \cite{Bakala2019}
\begin{eqnarray}
&&a(n)^{\hat \theta}+k_{\rm (Lie)}(n)^{\hat \theta}\,\nu^2+2\nu\sin^2\psi\ \theta(n)^{\hat \theta}{}_{\hat \varphi}\label{eq:susporbit}\\
&&+\frac{A (1+bN^\varphi)^2(1-\cos^2\beta\sin\psi)\cos\beta}{\gamma N^2\sqrt{\left( r^{2} + a^{2}-ab \right)^{2}- \Delta \left[ q + \left( b - a \right) ^{2} \right]}\tan\psi}=0,\notag
\end{eqnarray}
which permits to be solved in terms of $\psi$. Instead in the case of $b=0$ we obtain $\psi=\pm\pi/2$ \cite{DeFalco20183D,Bakala2019}. The critical points are either the suspended orbits or the equatorial ring, where the test particle ends its motion. In the Schwarzschild case, Eq. (\ref{eq:susporbit}) is an identity, because the test particle either stops on a point (for $b=0$) or move on a purely circular orbit in the equatorial ring (for $b\neq0$) of the critical hypersurface (see Refs. \cite{Bini2009,Bini2011,DeFalco20183D,Bakala2019}, for details).

The test particle comoves with the local corototaing observer frames in bound quasicircular orbits and in the equatorial plane in circular orbits \cite{Bini2011,DeFalco20183D}. In the Kerr case, the critical hypersurface assumes a quasi-ellipsoid shape, depending on the radiation emitting source parameters $(R_\star,\Omega_\star,A)$. Therefore, the critical radius is a function of $r_{\rm crit}=r_{\rm crit}(\theta,A,R_\star,\Omega_\star)$. In the Schwarzschild case instead, the critical hypersurface becomes a sphere, where such radius can be easily computed at the equator through Eq. (2.33) of Ref. \cite{Bini2011}. However, it is important to note that the non-vanishing angular velocity of the emitting surface $\Omega_\star\neq0$ breaks the spherical symmetry of the Schwarzschild metric \cite{Bakala2019}, so the 2D model is not anymore valid,  and only the 3D case must be employed.

\begin{figure*}[t!]
	\centering
	\hbox{\hspace{0cm}
		\includegraphics[scale=0.26]{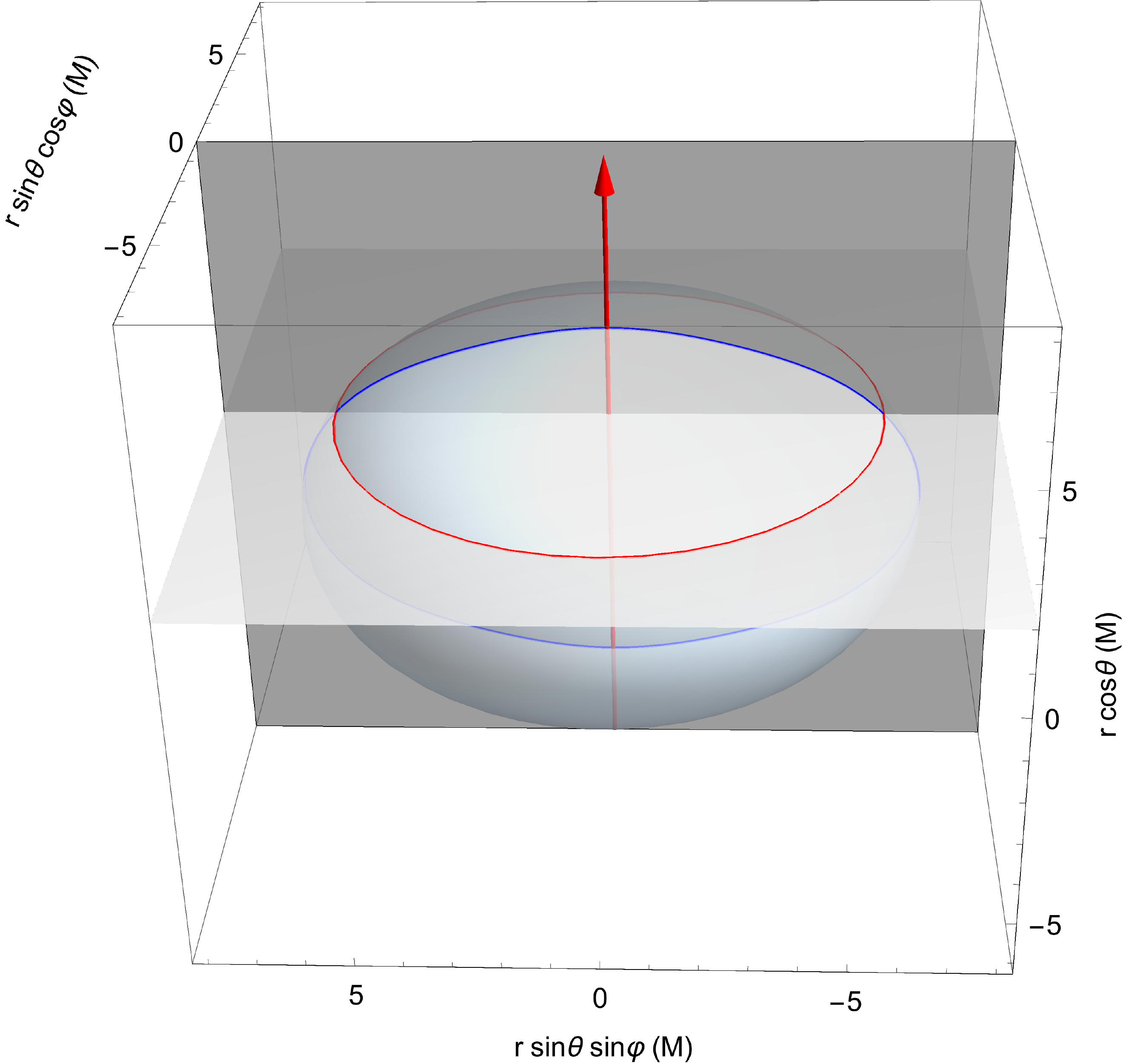}
		\hspace{0.4cm}
		\includegraphics[scale=0.49]{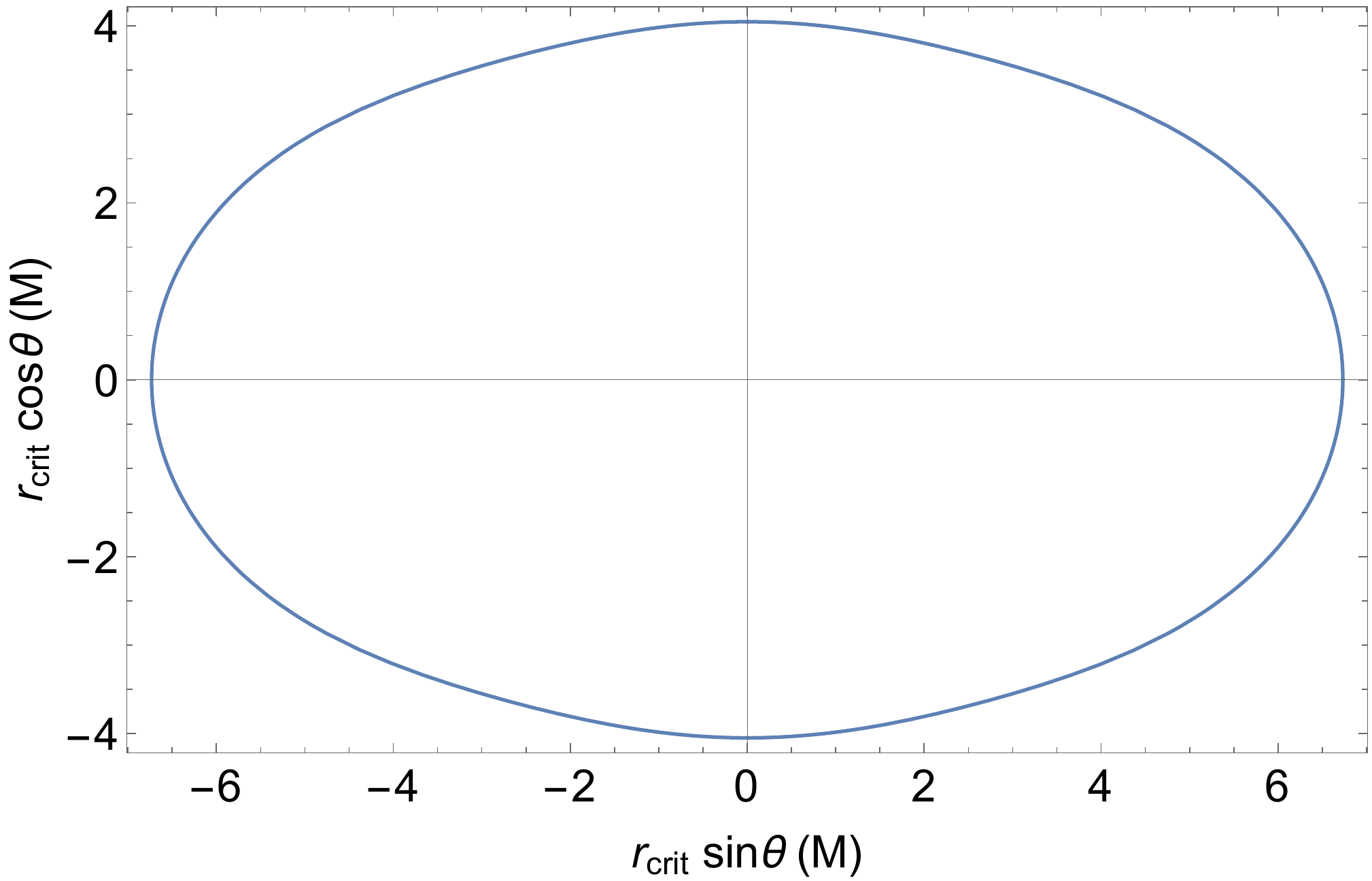}}
	\caption{Critical hypersurface around a NS of mass $M=1.5M_\odot$, rotating with frequency $600$ Hz (corresponding to an angular frequency of $\Omega_\star=0.031$ and spin of $a=0.41$), emitting spherical source of radius $R_\star=6M$, and luminosity parameter $A=0.7$. 	Left panel: 3D plot of the critical hypersurface intersected by the planes of equation $x=0$ and $z=2$. The red arrow is the polar $z$-axis, the blue and red curves are the intersections between the plane $x=0$ and $z=2$ and the critical hypersurface, respectively. The polar intersection angle is $\theta_{\rm int}=71.3^\circ$, the corresponding critical radius is $r_{\rm crit}(\theta_{\rm int})=6.23M$, and the radius of the red ring is $r_{\rm crit}(\theta_{\rm int})\sin\theta_{\rm int}=5.9M$. Right panel: 2D plot of the blue curve, where it is possible to note the symmetry with respect to the equatorial and polar planes (azimuthal rotational symmetry).}
	\label{fig:Fig1}
\end{figure*}

In addition, there are also the conditions $dr/d\tau=d\theta/d\tau=0$ (based on the critical hypersurface definition). This means that at fixed radius $r$, the polar angle $\theta$ is constant ($d\theta/d\tau=0$), while at fixed polar angle $\theta$ the radius $r$ is constant ($dr/d\tau=0$). In other words, such conditions require that each parallel section to the equatorial plane intersecting the critical hypersurface gives a circular ring. For the azimuthally rotational simmetry, the critical hypersurfaces are surfaces of revolution symmetric with respect to the polar axis (see Fig. \ref{fig:Fig1}).

\section{Stability of the critical hypersurfaces}
\label{sec:stability}
For performing our proof, we set before our notations. The dynamical system (\ref{EoM1})--(\ref{EoM5}) is represented by\footnote{We do not include Eq. (\ref{EoM6}) because it is not vanishing at the critical hypersurface. In addition, being the dynamical system symmetric with respect to $\varphi$-rotations, it is possible to remove such equation without any lost of generality. Equation (\ref{time}) is also not added, because it expresses only the conversion between proper time $\tau$ and coordinate time $t$.}
\begin{equation}\label{eq:DS}
\dot{\boldsymbol{x}}=\boldsymbol{f}(\boldsymbol{x}),
\end{equation}
where $\dot{\boldsymbol{x}}=d\boldsymbol{x}/d\tau$ is defined in $\mathcal{D}$, representing the spatial region outside the compact object together with the admissible velocity field field of the test particle, namely
\begin{equation}
\begin{aligned}
\mathcal{D}&=\left\{\nu\in[0,1],\ \psi\in[0,\pi],\ \alpha\in[0,2\pi],\right.\\ 
&\left. \qquad r\in]2M,\infty[,\ \theta\in[0,\pi]\right\}.
\end{aligned}
\end{equation}
A point $\boldsymbol{x_0}=(\nu_0,\psi_0,\alpha_0,r_0,\theta_0)$ is called \emph{critical}, if it vanishes Eq. (\ref{eq:DS}), i.e., $\boldsymbol{f}(\boldsymbol{x_0})=\boldsymbol{0}$. The set of all critical points of Eq. (\ref{eq:DS}) is called \emph{critical hypersurface}, namely 
\begin{equation}
\mathcal{H}=\left\{\boldsymbol{x_0}\in\mathcal{D}\ |\ \boldsymbol{f}(\boldsymbol{x_0})=\boldsymbol{0}\right\}\subseteq \mathcal{D}.
\end{equation}
The function $\Phi:\mathcal{D}\times\mathbb{R}\longrightarrow\mathcal{D}$ is called the \emph{flow} associated to the dynamical system (\ref{eq:DS}). Called $\boldsymbol{x_i}$ the initial conditions, we have that the solution at time $\bar{\tau}$, indicated by $\boldsymbol{x}(\bar{\tau})$, can be also written as $\boldsymbol{x}(\bar{\tau})=\Phi_{\bar{\tau}}(\boldsymbol{x_i})\equiv\Phi(\bar{\tau},\boldsymbol{x_i})$.

\begin{figure*}[t!]
	\centering
	\hbox{\hspace{0cm}
		\includegraphics[trim=1.9cm 1cm 1cm 2cm,scale=0.365]{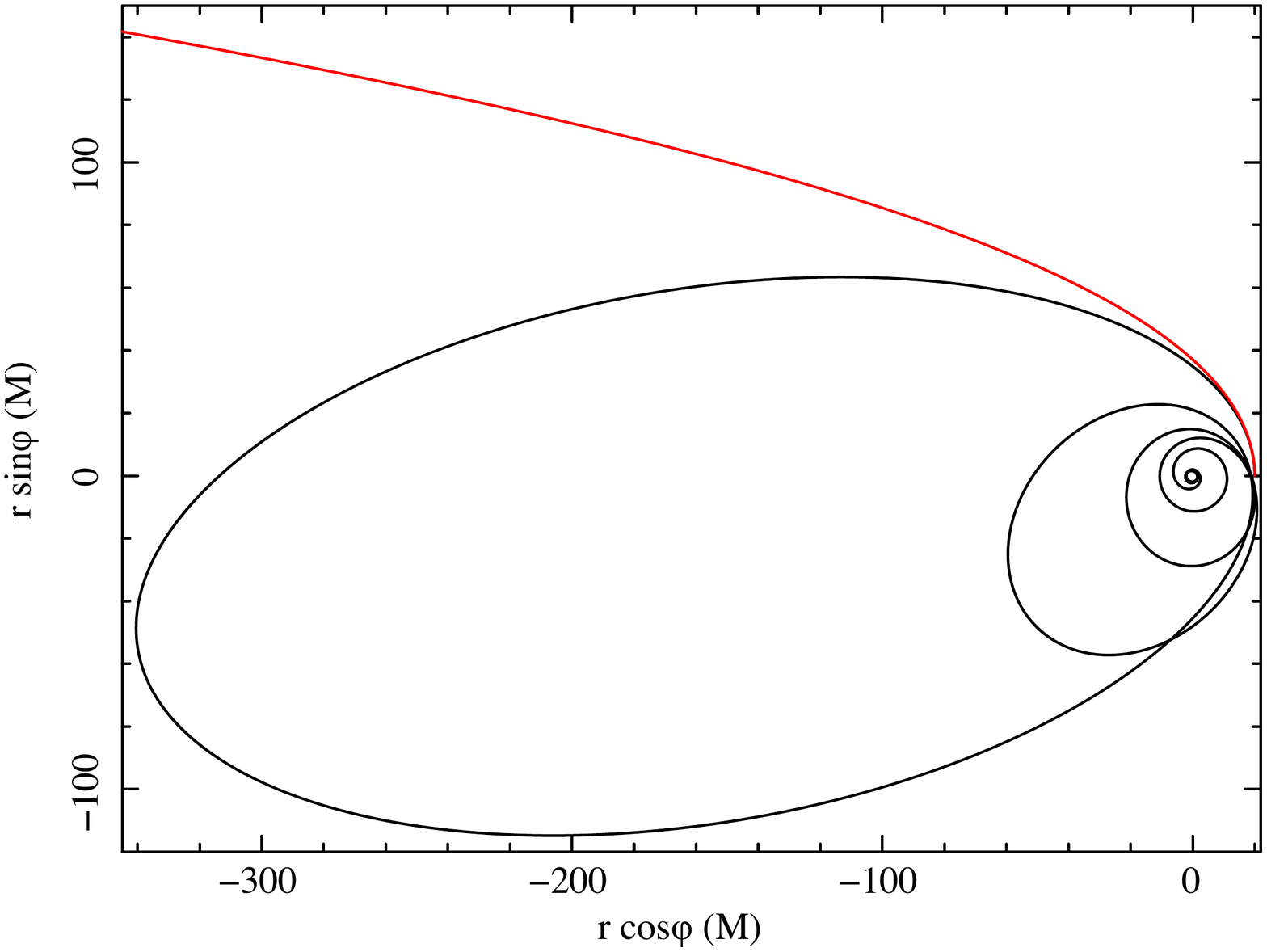}
		\includegraphics[trim=1.9cm 1cm 1cm 2cm,scale=0.365]{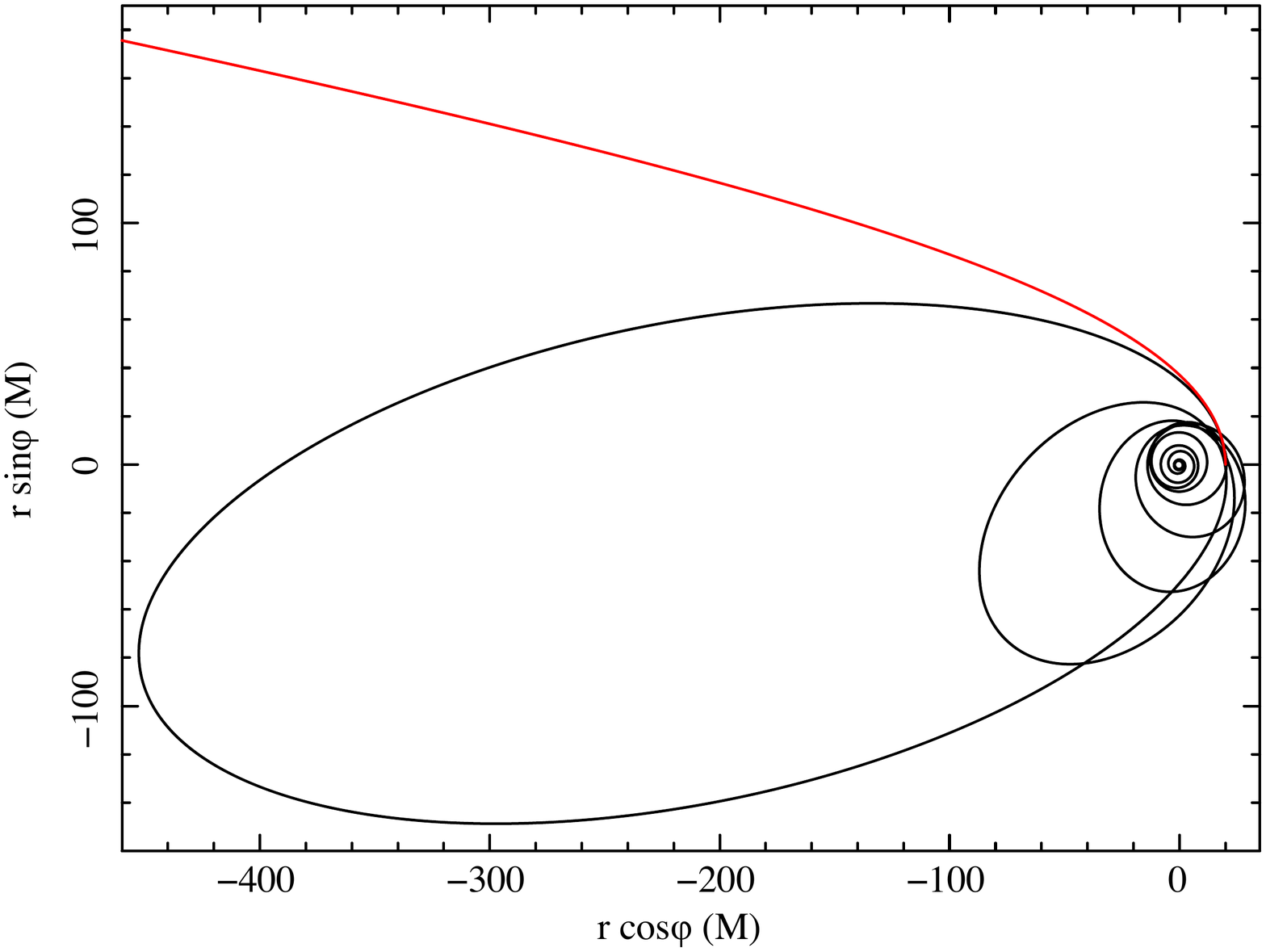}}
	\caption{Motions of test particles in the equatorial plane of a Kerr BH of spin $a=0.4$ and luminosity parameter $A=0.1$. The test particles in both panels have initial positions $(r_0,\varphi_0)=(20M,0)$ and initial velocities $(\nu_0,\alpha_0)=(0.3,0)$ for black orbits, and $(\nu_0,\alpha_0)=(0.31,0)$ for red orbits. The photon impact parameters are $b=0$ (left panel) and $b=2$ (right panel).}
	\label{fig:Fig2}
\end{figure*}

We consider only those initial configurations, where the test particle ends its motion on the critical hypersurface without escaping at infinity. Unfortunately, it is not simple to mathematically characterize such class of solutions, because the dynamical system under study is too complex. In addition, there is a strong dependence not only on the input parameters determining the radiation field, but also on the test particle initial data. Indeed, this dynamical system shows a \emph{sensitive dependence on the initial conditions}, whose propriety is to exhibit extremely different behaviors with only tiny changes in the initial conditions\footnote{In Ref. \cite{DeFalco20183D}, it is explicitly stated that the 3D case is more sensitive from the initial conditions with respect to the 2D case. Indeed, to solve such issue there are more controls on the integration process reaching thus an average relative accuracy of $\sim10^{-14}$.} (see Fig. \ref{fig:Fig2}, as example). 

Once the stability has been proven, it immediately follows that the critical equatorial ring is a \emph{stable attractor}, i.e., a region toward which the test particle tends to be attracted and to end its motion. More formally, a set $\mathcal{A}$ is an attractor for the dynamical system (\ref{eq:DS}) if \cite{Milnor1985,Robert2004}
\begin{enumerate}
\item $\mathcal{A}$ is \emph{forward invariant} under $\Phi_\tau$, namely if $\boldsymbol{x_0}\in\mathcal{A}$ then also $\Phi_\tau(\boldsymbol{x_0})\in\mathcal{A}$ for all $\tau > 0$;
\item there exists a neighborhood of $\mathcal{A}$, called the \emph{basin of attraction for $\mathcal{A}$}, denoted by $\mathcal{B}(\mathcal{A})$, which consists of all points $\boldsymbol{x}$ that enter $\mathcal{A}$ in the limit $\tau\to\infty$. More formally, $\mathcal{B}(\mathcal{A})$ is the set of all points $\boldsymbol{x}$ in the phase space with the following property: for any open neighborhood $\mathcal{U}$ of $\mathcal{A}$, there is a positive constant $\mathcal{T}$ such that $f(t,\boldsymbol{x})\in\mathcal{U}$ for all real $\tau > \mathcal{T}$;
\item there is no proper (non-empty) subset of $\mathcal{A}$ having the first two properties. In other words, $\mathcal{A}=\cap_{\tau\ge 0} \Phi_\tau(\mathcal{U})$, where $\mathcal{U}$ is a basin of attraction.
\end{enumerate}
The first and second propriety are basically linked to the proof of the stability of the critical hypersurface, while the third shows that the critical hypersurface $\mathcal{H}=\mathcal{A}$, because once the test particle moves on $\mathcal{H}$, it will not leave such region, configuring thus as the smallest basin of attraction for the dynamical system (\ref{eq:DS})\footnote{\label{footnote}It must be distinguished two cases depending on the initial conditions of the test particle position \cite{Bini2011,DeFalco20183D}. If the test particle is set outside the compact object and reaches the critical hypersurface, it will stay over there forever. Instead, if the test particle starts inside the critical hypersurface it could cross the critcal hypersurface once,  because the test particle has sufficient energy to be not attracted. Then, if it does not escape at infinity, it will fall on the critical hypersurface and will stay over there forever,  as it started its motion outside. Therefore for such reason, the test particle cannot cross the critical hypersurface more than once.}. Another important propriety of the critical hypersurfaces is they are compact sets (see Eq. (\ref{eq:FCR}), for more details).

In the next sections, we show how to formally prove the stability of the critical hypersurfaces by recalling what has been done in the literature (see Sec. \ref{sec:LST}) and then by introducing our new contributions (see Sec. \ref{sec:LF}).   

\subsection{Linear stability theory}
\label{sec:LST}
Bini and collaborators \cite{Bini2009,Bini2011} have previously presented the proof of the stability by performing calculations within the \emph{linear stability theory} \cite{Robert2004}. This method consists in linearizing the non-linear dynamical system towards the critical points of the critical hypersurface, i.e.,
\begin{equation} \label{eq:DSL}
\dot{\boldsymbol{x}}\approx \boldsymbol{A}\cdot (\boldsymbol{x}-\boldsymbol{x_0}),
\end{equation}
where $\boldsymbol{A}=(\partial_{\boldsymbol{x}}\boldsymbol{f})(\boldsymbol{x_0})$ is a linear operator. 

Then, after having diagonalized the matrix $\boldsymbol{A}$, one looks at its eigenvalues and check whether they are negative or have real part negative for inferring the stability of the critical points. Such procedure holds whenever the critical points are not \emph{hyperbolic}, meaning that the matrix $\boldsymbol{A}$ has no eigenvalues with real part equal to zero. Indeed, the \emph{Hartman-Grobman theorem} (or also known in the literature as \emph{linearization theorem}) states that \cite{Hartman1964,Arrowsmith1992} there exists a neighborhood $\mathcal{U}$ of $\boldsymbol{x_0}$ and a homeomorphism $\boldsymbol{h}\equiv\boldsymbol{A}\cdot (\boldsymbol{x}-\boldsymbol{x_0}):\mathcal{U}\to\mathbb{R}^{n}$ with $\boldsymbol{h}(\boldsymbol{x_0})=\boldsymbol{0}$ such that in the neighborhood $\mathcal{U}$ the dynamical system (\ref{eq:DS}) is topologically homeomorphic to the dynamical system (\ref{eq:DSL}) through the map $\boldsymbol{h}$. 

For the difficulty of calculations Bini and collaborators have only shown the stability of the 2D critical hypersurfaces in the Schwarzschild case (see Appendix in Ref. \cite{Bini2011}). This method, albeit simple in its theoretical explanation, practically requires to develop several calculations (especially in the Kerr case), therefore we are looking for a new, simpler, and more physical approach.

\subsection{Lyapunov function}
\label{sec:LF}
We propose a new method framed within the \emph{Lyapunov theory} \cite{Robert2004}, which is easier both in terms of calculations and gives more physical insight into the problem under investigation. Let $\Lambda=\Lambda(\nu,\psi,\alpha,r,\theta)$ be a real valued function of the test particle position and velocity fields, continuously differentiable in all points outside of the compact object. Then $\Lambda$ is a \emph{Lyapunov function} for the set $\mathcal{H}$, if it fulfills the following conditions:
\begin{eqnarray}  
{\rm (I)}&&\quad \Lambda(\boldsymbol{x})>0,\quad \forall \boldsymbol{x}\in \mathcal{D}\setminus\mathcal{H};\label{eq:lia1}\\
{\rm (II)}&&\quad \Lambda(\boldsymbol{x_0})=0,\quad \forall \boldsymbol{x_0}\in \mathcal{H};\label{eq:lia2}\\ 
{\rm (III)}&&\quad \dot{\Lambda}(\boldsymbol{x})\equiv\nabla\Lambda(\boldsymbol{x})\cdot \boldsymbol{f}(\boldsymbol{x})\le0 ,\quad \forall \boldsymbol{x}\in \mathcal{D}\label{eq:lia3}.
\end{eqnarray}
Once the Lyapunov function $\Lambda$ has been found for all points belonging to the critical hypersurface $\mathcal{H}$, a theorem due to Lyapunov assures that $\mathcal{H}$ is stable \cite{Hahn1967,Robert2004}. In addition, if the third condition (\ref{eq:lia1}) is replaced by
\begin{equation} \label{eq:lia3}
{\rm(III')}\quad \dot{\Lambda}(\boldsymbol{x})<0 ,\quad \forall \boldsymbol{x}\in \mathcal{D},
\end{equation}
then $\mathcal{H}$ is asymptotically stable \cite{Hahn1967,Robert2004}. 

The great advantage of such a method relies on the fact that it can be applied without solving the differential equations (\ref{eq:DS}). It is important to note that the Lyapunov function is not unique at all, indeed there could be the cases where it is possible to find just one, more than one, or even anything. Unfortunately, there is no a mathematical recipe for determining a Lyapunov function, because it is usually a matter of ingenuity, trials, or luck in each case. However, sometimes there are natural functions to try, like for example the associated first integrals (see Ref. \cite{Robert2004}, for examples). In our case we were able to determine three different Lyapunov functions, having all important physical meanings. 

\subsubsection{Energy}
\label{sec:KE}
We propose as first Lyapunov function the  test particle relative energy, where both kinetic and potential energies related to the radiation pressure and the gravitational force are all measured in the ZAMO frame, i.e., 
\begin{equation} \label{eq:LF1}
\mathbb{K}=\frac{m}{2}\left|\nu^2-\nu^2_{\rm crit}\right|+(A-M)\left(\frac{1}{r}-\frac{1}{r_{\rm crit}}\right),
\end{equation}
where $m$ is the the test particle mass and $\nu_{\rm crit}(\theta)=[\cos\beta]_{r=r_{\rm crit}(\theta)}$, which includes as a particular case the velocity in the equatorial ring $\nu_{\rm eq}=[\cos\beta]_{r=r_{\rm crit}(\pi/2)}$. 

By definition it is defined positive outside the critical hypersurface, because the second term is the product of two negative terms (since $A/M \in [0,1]$ and $r\ge r_{\rm crit}$), satisfying thus condition (I), and it is identically zero on the critical hypersurface, as requested by condition (II). Regarding the condition (III) we need to calculate the $\tau$-derivative of $\mathbb{K}$, which is given by
\begin{equation}  \label{eq:DLF1}
\begin{aligned}
\dot{\mathbb{K}}&=m\ {\rm sgn}\left(\nu^2-\cos^2\beta\right)\left[\nu\frac{d\nu}{d\tau}-\cos\beta\frac{d (\cos\beta)}{d \tau}\right]\\
&\quad-\frac{A-M}{r^2}\dot{r}.
\end{aligned}
\end{equation}
where ${\rm sgn}(x)$ is the signum function. In Appendix \ref{sec:AKE} we prove that $\dot{\mathbb{K}}$ is definite non-positive. In Fig. \ref{fig:Fig3} we show an example of test particle trajectory and the related mechanical energy, together with its $\tau$-derivative (see upper right panel). As we can graphically see, $\mathbb{K}$ fulfills all the conditions to be a Lyapunov function. 

\subsubsection{Angular momentum}
\label{sec:AM}
Now, we consider as second Lyapunov function the relative angular momentum of the test particle measured in the ZAMO frame, i.e., 
\begin{equation}  \label{eq:LF2}
\begin{aligned}
\mathbb{L}=m(r\nu\sin\psi\cos\alpha-r_{\rm crit}\nu_{\rm crit}).
\end{aligned}
\end{equation}
By definition it is defined positive outside the critical hypersurface, satisfying thus condition (I), and it is identically zero on the critical hypersurface, as requested by condition (II). Regarding the condition (III) we need to calculate the $\tau$-derivative of $\mathbb{L}$, which is given by
\begin{eqnarray}  \label{eq:DLF2}
\dot{\mathbb{L}}&&=m\left[-\dot{r}_{\rm crit}\nu_{\rm crit}-r_{\rm crit}\frac{d(\nu_{\rm crit})}{d\tau}+r\frac{d\nu}{d\tau}\cos\alpha\sin\psi\right.\\
&&\left.\nu(\dot{r}\cos\alpha\sin\psi-r\sin\alpha\sin\psi\ \dot{\alpha}+r\sin\alpha\cos\psi\ \dot{\psi})\right].\notag
\end{eqnarray}
In Appendix \ref{sec:AAM} we prove that $\dot{\mathbb{L}}$ is definite non-positive. In Fig. \ref{fig:Fig3} we show the  angular momentum and its $\tau$-derivative behaviors (see lower left panel). Therefore, graphically we see that also the function $\mathbb{L}$ fully respects the conditions to be a Lyapunov function.

\subsubsection{Rayleigh potential}
\label{sec:RP}
In the study of the general relativistic PR effect, it has been proved it admits a Lagrangian formulation, albeit it is a highly-non-linear dissipative system in GR \cite{DeFalco2018}. This can be realised through the help of an integrating factor $\mu=E_p^2/\mathbb{E}^2$ \cite{DeFalco2019,DeFalco2019VE}, where $E_p$ is the photon energy and $\mathbb{E}\equiv E(U)=-k_\alpha U^\alpha$, see Eq. (\ref{enepart}). 

Using the \emph{energy formalism} \cite{DeFalco2019} it was possible to determine the explicit formula for the Rayleigh potential $\mathbb{F}$ related to the radiation force $\mu {\mathcal F}_{\rm (rad)}(U)^\alpha$ (see Eq. (\ref{radforce})). Therefore, the third Lyapunov function is the relative Rayleigh potential\footnote{The Rayleigh pitential is composed by an absorption and emission term. However, it is possible to see that the emission term is a constant since it is equal to the module of the test particle velocity, i.e. $U_\alpha U^\alpha=-1$ (see Ref. \cite{DeFalco2019,DeFalco2019VE}, for more details)}, i.e.,
\begin{equation} \label{eq:LF3}
\mathbb{F}=\tilde{\sigma}\mathcal{I}^2\left[\lg\left(\frac{\mathbb{E_{\rm crit}}}{E_p}\right)-\lg\left(\frac{\mathbb{E}}{E_p}\right)\right],
\end{equation}
where $\mathbb{E_{\rm crit}}$ is the energy $\mathbb{E}$ evaluated on the critical hypersurface, and it is simply given by 
\begin{equation} 
\begin{aligned}
\mathbb{E_{\rm crit}}&=[\mathbb{E}]_{r=R_\star,\alpha=0,\pi,\psi=\pm\pi/2,\nu=\nu_{\rm crit}}\\
&=\frac{E_p|(\sin\beta)_{\rm crit}|}{N_{\rm crit}}(1+bN^\varphi_{\rm crit}),
\end{aligned}
\end{equation}
where the subscript \qm{crit} means to evaluate a quantity at the critical hypersurface.

\begin{figure*}[t!]
	\centering
	\vbox{
	\hbox{\hspace{0cm}
		\includegraphics[scale=0.345]{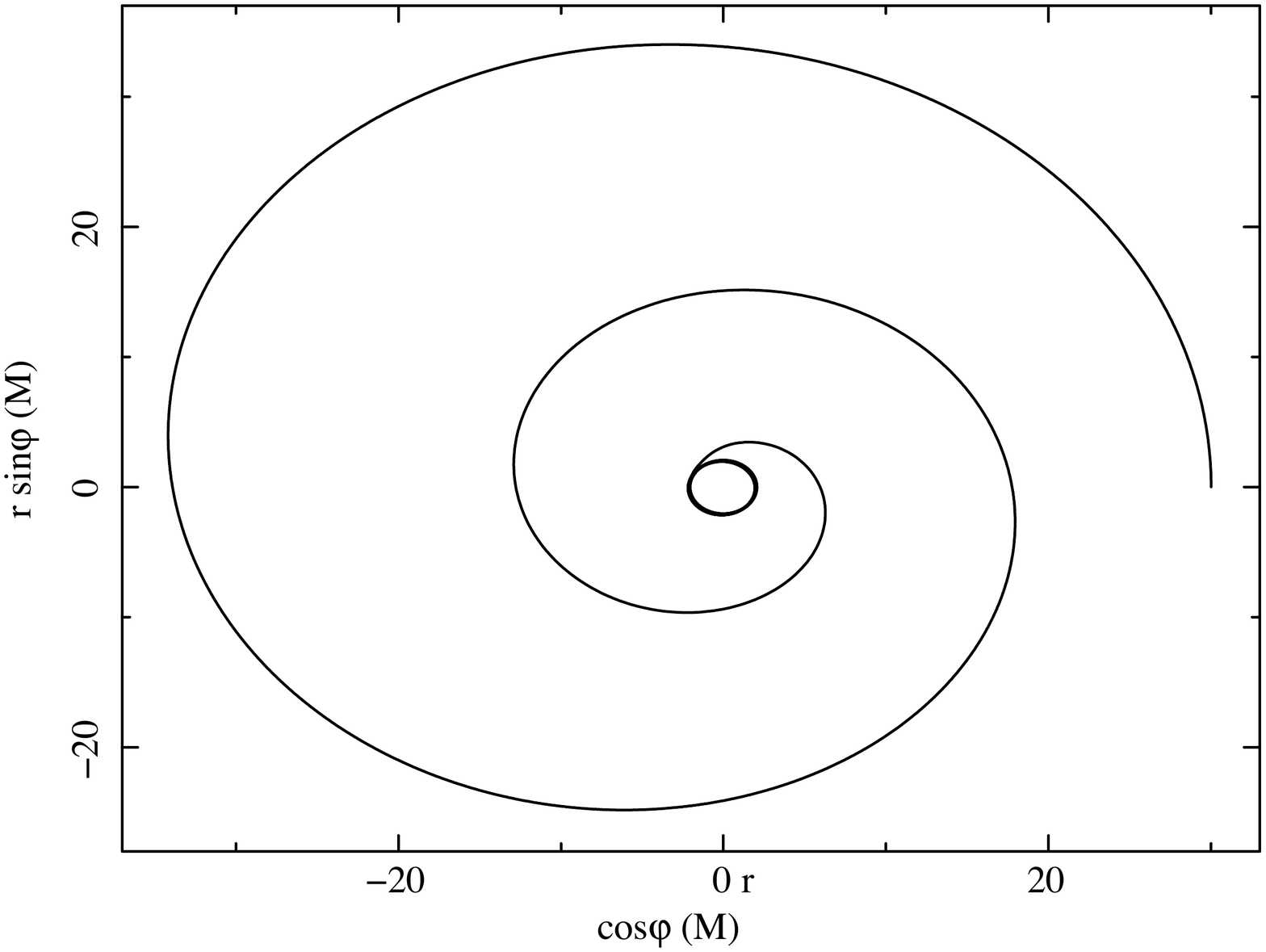}
		\hspace{-1 cm}
		\includegraphics[scale=0.345]{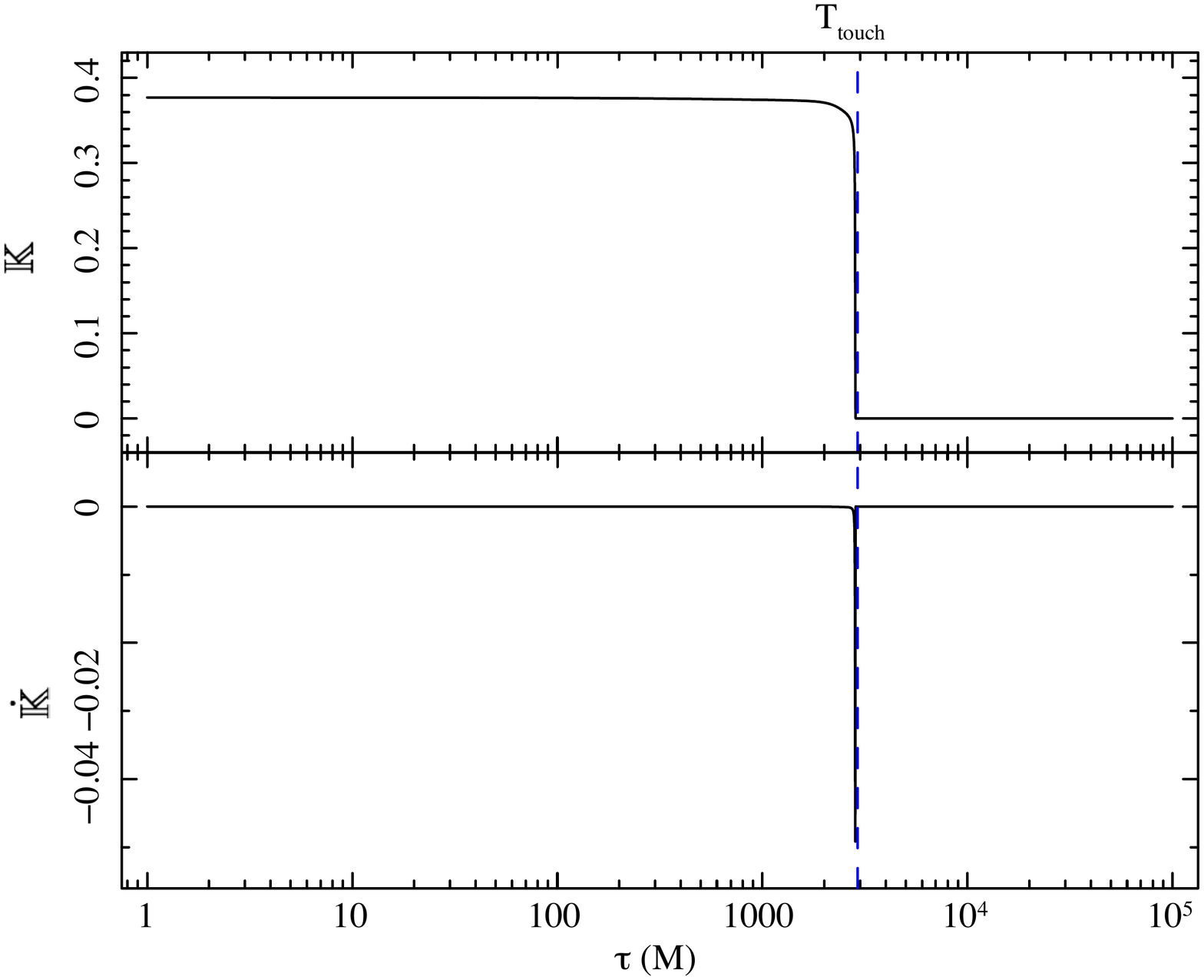}}
		\vspace{-0.3 cm}
	\hbox{\hspace{0cm}
		\includegraphics[scale=0.345]{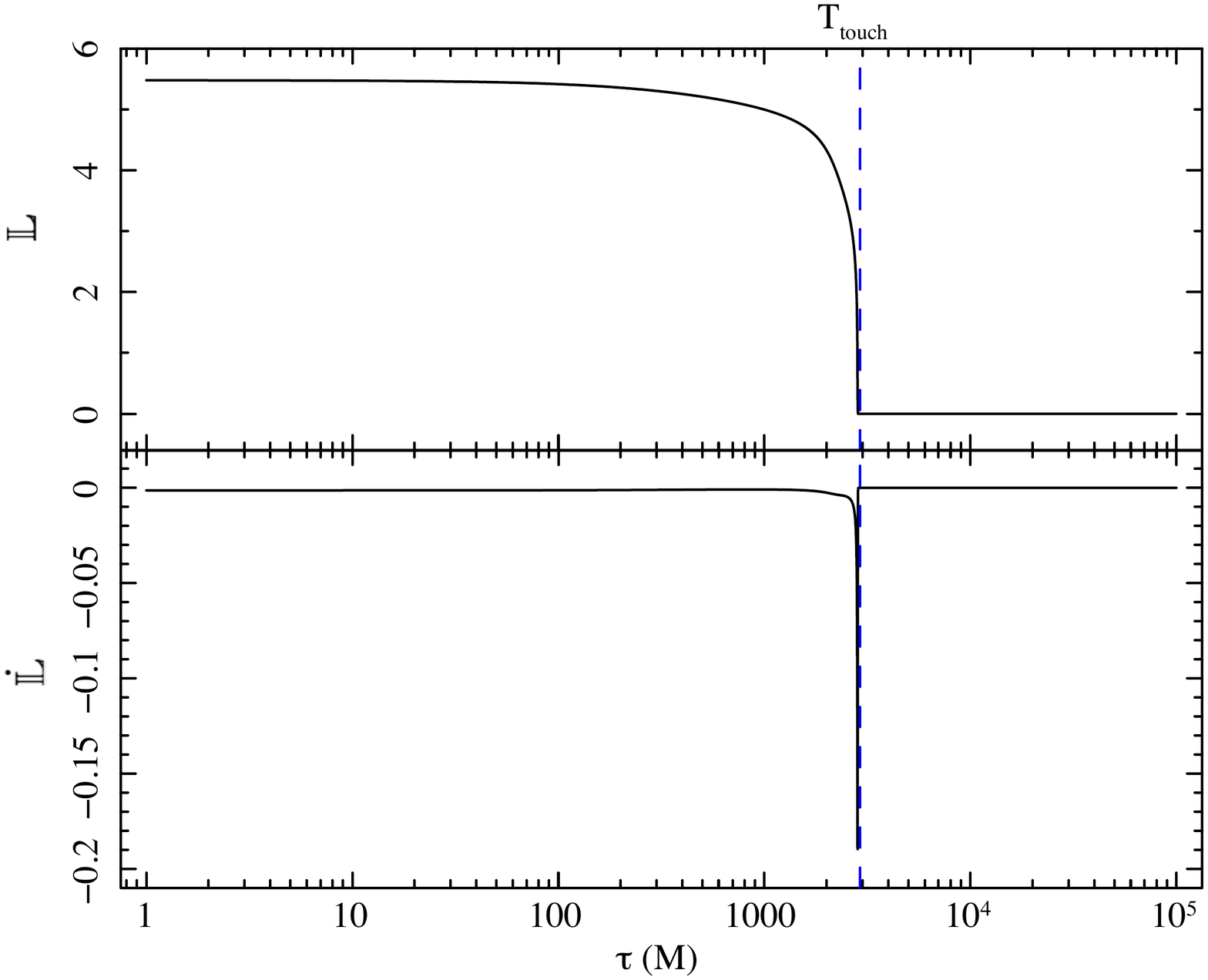}
		\hspace{-1 cm}
		\includegraphics[scale=0.345]{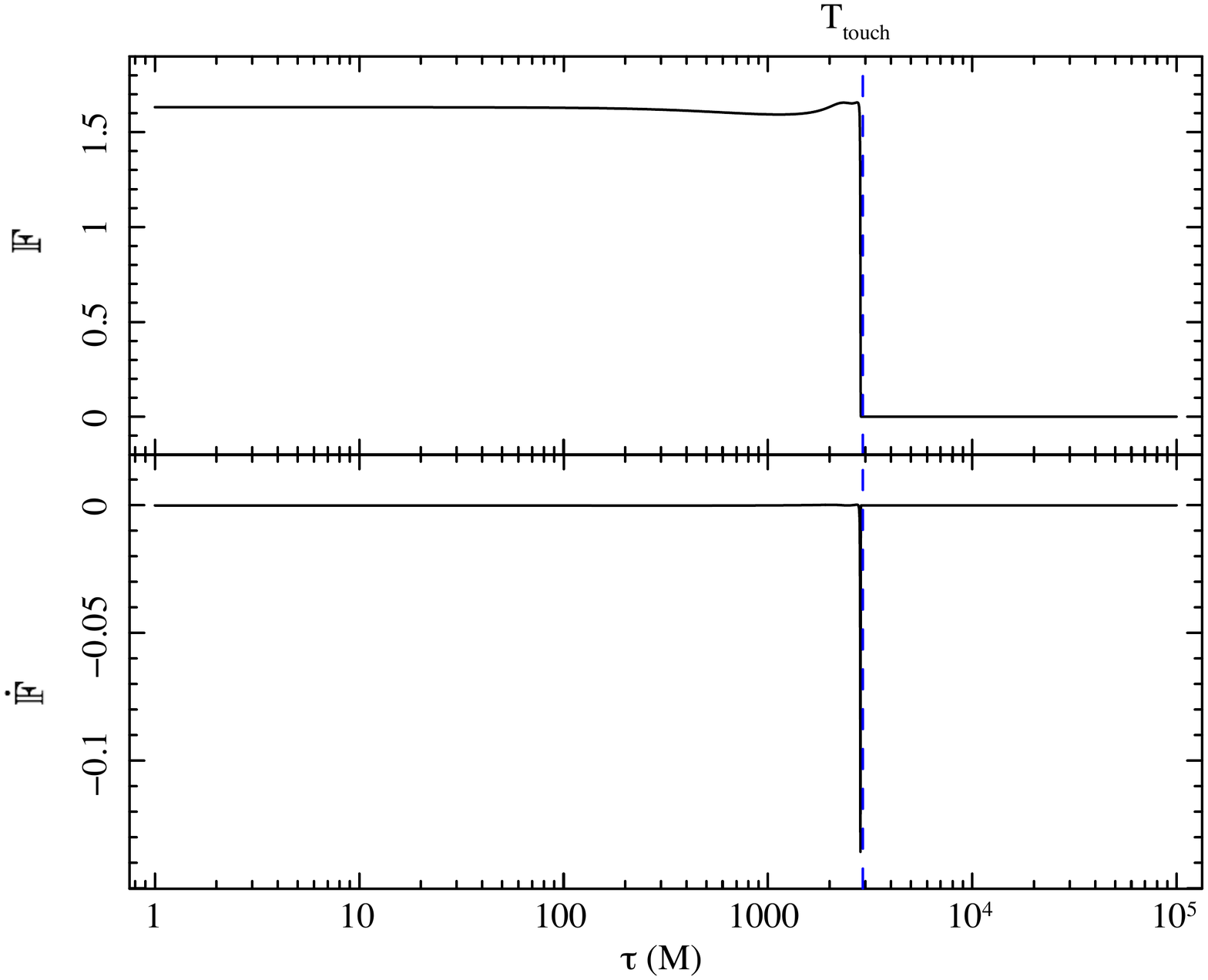}}}
	\caption{We show a test particle orbit and the related three Lyapunov functions, which graphically prove the stability of the critical hypersurface. \emph{Upper left panel:} test particle moving around a rotating compact object with mass $M=1$, spin $a=0.3$, luminosity parameter $A=0.2$, and photon impact parameter $b=0$. The test particle starts its motion at the position $(r_0,\varphi_0)=(30M,0)$ with velocity $(\nu_0,\alpha_0)=(\sqrt{M/r_0},0)$. The critical hypersurface is a circle with radius $r_{\rm crit}=2.07M$. The energy (see Eqs. (\ref{eq:LF1}) and (\ref{eq:DLF1}), and \emph{upper right panel}), the angular momentum  (see Eqs. (\ref{eq:LF2}) and (\ref{eq:DLF2}), and \emph{lower left panel}), and the Rayleigh potential (see Eqs. (\ref{eq:LF3}) and (\ref{eq:DLF3}), and \emph{lower right panel}) together with their $\tau$-derivatives are all expressed in terms of the proper time $\tau$. The dashed blue lines in all plots represent the proper time $T_{\rm touch}$ at which the test particle reaches the critical hypersurface and it amounts to $T_{\rm touch}=2915M$. }
	\label{fig:Fig3}
\end{figure*}

The energy absorbed by the test particle $\mathbb{E}$ is always minor than the photon energy $E_p$, and its maximum is attained at $E_p$ only when the test particle is in rest \cite{DeFalco2019}. Therefore, it is everywhere positive, satisfying thus the condition (I), and the condition (II) on the critical hypersurface. The $\tau$-derivative of $\mathbb{F}$ is 
\begin{equation}  \label{eq:DLF3}
\begin{aligned}
\dot{\mathbb{F}}&=\tilde{\sigma}\dot{(\mathcal{I}^2)}\left[\lg\left(\frac{\mathbb{E_{\rm crit}}}{E_p}\right)-\lg\left(\frac{\mathbb{E}}{E_p}\right)\right] \\
&+\tilde{\sigma}\mathcal{I}^2\left[\frac{\dot{\mathbb{E}}_{\rm crit}}{\mathbb{E_{\rm crit}}}-\frac{\dot{\mathbb{E}}}{\mathbb{E}}\right].
\end{aligned}
\end{equation}
In Appendix \ref{sec:ARP} we prove that $\dot{\mathcal{F}}$ is definite non-positive, verifying thus condition (III). In Fig. \ref{fig:Fig3} we show how the profile of the Rayleigh potential respects the conditions to be a Lyapunov function (see lower right panel). 

\section{Conclusions}
\label{sec:end}
In this paper, we have shown how to formally prove the stability of the critical hypersurfaces for the general relativistic (both 2D and 3D) PR effect models. We have proposed a new approach based on the Lyapunov functions, which is more elegant, easier in terms of calculations, and also contains important physical information on the system under study. Previously in the literature, it has been exploited the linearized theory around the critical points, but it revealed to be not so powerful, because it requires strong computational efforts in linearizing the dynamical system and then in finding the eigenvalues and studying the sign, especially for the 3D case. For such reasons, only the Schwarzschild (2D) case has been completely proved, since in the Kerr spacetime everything becomes extremely more difficult to handle \cite{Bini2011}. 

Our method is able to fully prove the stability issue, without recurring to any symbolic program or dedicated software to carry out our analysis. All the estimations and calculations reported in this paper can be relatively easily handled. The Lyapunov theory applied to our problem reveals to be very ingenious, clear in the calculation process, and more powerful than a numerical-programming approach. In addition, we understand the contribution played by each single term present in the dynamical equations (see Appendices \ref{sec:AKE}, \ref{sec:AAM}, \ref{sec:ARP}). 

In addition, we were able to find and propose three different Lyapunov functions with a well precise meaning, proving thus the stability in different ways. Since the PR effect removes energy and angular momentum from the test particle, those were the inspiring ideas, which led us to build up the first two Lyapunov functions (see Secs. \ref{sec:KE} and \ref{sec:AM}). They represent the classical version and not the general relativistic expression. Even if we used not the proper definition, they permit to easily carry out the calculations and achieve the stability results. There is no contradictions with the definition of Lyapunov function and its application. We note that even a mathematical function, with no physical meaning connected with the system under study, but verifying the Lyapunov conditions would be a good candidate for proving the stability of the critical hypersurfaces.      

The third Lyapunov function is less intuitive than the previous cases, because it stems out from the Lagrangian approach to the general relativistic PR effect \cite{DeFalco2018,DeFalco2019,DeFalco2019VE}. Indeed, employing the \emph{energy formalism}, it is possible to analytically derive the general relativistic Rayleigh potential, which contains the radiation field absorption processes affecting the test particle motion. Such a function involves the logarithm of the test particle absorbed energy. We thought about this function by looking at the plots and physical meaning reported in Refs. \cite{DeFalco2019VE}. Through these valuable results, we have understood that the critical hypersurfaces are basin of attraction and the equatorial ring is a stable attractor. 


This approach shows also another great potentiality, because it can be exploited to prove the stability of the critical hypersurfaces in further possible extensions of the general relativistic PR effect model, naturally with the due modifications, and still keeping its good performances. In a future work, we would like to deeply investigate the proprieties of this effect under the dynamical system point of view. Such kind of research, which has never been studied or proposed so far in the literature, will shed new light on the PR effect. The difficulties of such approach rely on mathematically formalising some notions used intuitively so far. These efforts will permit to develop new methods and techniques apt to infer not only new interesting results on the PR effect, but more in general on dissipative systems in GR.

\section*{Acknowledgements}
V.D.F. thanks the Silesian University in Opava for having funded this work. V.D.F. and P.B. acknowledge the Osservatorio Astronomico di Roma in Monteporzio Catone for the hospitality, since part of this work has been carried out there. V.D.F. is grateful to Gruppo Nazionale di Fisica Matematica of Istituto Nazionale di Alta Matematica for support. P.B. acknowledges the Czech Science Foundation (GAČR) grant GAČR 17-16287S and internal grant of Silesian University in Opava SGS/13/2019. V.D.F. and P.B. are grateful to Prof. Luigi Stella for the useful discussions.

\begin{appendix}
\section{General remarks}
Before starting with our analysis, it is useful to estimate the order of magnitude of all metric and ZAMO kinematical quantities. Since we consider test particle orbits ending their motions on the critical hypersurface, it is reasonable to assume that the test particle is confined in a box during its evolution, which means to impose the following reasonable limits on the test particle position and velocity parameters\footnote{We prove the stability for test particle set outside the critical hypersurface. The proof can be easily extended also to a test particle set inside the critical hypersurface (but outside the emitting surface), see also the argument of footnote \ref{footnote}.}:
\begin{equation} \label{eq:cond}
\begin{aligned}
&r\in [R_\star,\bar{R}],\quad \varphi\in[0,2\pi],\quad\ \ \ \theta\in\left[0,\frac{\pi}{2}\right],\\
&\nu\in[0,1],\qquad \alpha\in\left[0,-\pi\right],\quad\ \psi\in\left[0,\frac{\pi}{2}\right], 
\end{aligned}
\end{equation}
where $\bar{R}<\infty$ is the maximum radial extension the test particle can reach, changing for each initial configuration. In addition, we note that $R_\star\ge r_{\rm crit}>R_H>M\ge1$, where $R_H$ is the event horizon radius. Considering $M\ge1$ is not a restrictive condition at all, because we can set the unity measure mass equal to one. 

Therefore, we have the following estimations (see Table 1 in Ref. \cite{DeFalco20183D}, for the full explicit expressions):
\begin{eqnarray} \label{eq:EST1} 
&&b\sim R_\star^2\Omega\sin^2\theta,\quad N\sim 1,\quad N^\varphi\sim-\frac{2Ma}{r^3},\notag\\
&&\Delta\sim r^2,\quad \rho\sim r^2,\quad \Sigma\sim r^2,\notag\\
&&\sqrt{g_{rr}}\sim 1,\quad \sqrt{g_{\theta\theta}}\sim r,\quad \sqrt{g_{\varphi\varphi}}\sim r|\sin\theta|,\notag\\
&&a(n)^{\hat r}\sim\frac{M}{r^2},\quad \theta(n)^{\hat r}{}_{\hat\varphi}\sim -\frac{aM\sin\theta}{r^3}, \\
&&a(n)^{\hat \theta}\sim-\frac{a^2M\sin\theta}{r^4},\quad \theta(n)^{\hat \theta}{}_{\hat\varphi}\sim -\frac{a^2M\sin\theta\sin2\theta}{r^5},\notag\\
&&k_{\rm (Lie)}(n)^{\hat r}\sim-\frac{1}{r},\quad k_{\rm (Lie)}(n)^{\hat \theta}\sim-\frac{\cot\theta}{r}.\notag
\end{eqnarray}

We estimate also the dynamical equations (\ref{EoM1})--(\ref{EoM5})
\begin{eqnarray} \label{eq:EST2}
&&\frac{d\nu}{d\tau}\sim\frac{M}{r^2}\left\{-\frac{\sin\alpha\sin\psi}{\gamma}\right.\notag\\
&&\left.+\frac{A}{M}[1-\nu\sin\psi\cos(\alpha-\beta)]\ [\sin\psi\cos(\alpha-\beta)-\nu]\right\}\notag\\
&&\frac{d\psi}{d\tau}\sim\frac{\gamma\nu}{r}\left(\sin\alpha\cos\psi-\cot\theta\cos^2\alpha\sin\psi\right),\notag\\
&&\frac{d\alpha}{d\tau}\sim\frac{\gamma \nu\cos\alpha}{r\sin\psi}\left[1+\cot\theta\cos^2\psi\sin\alpha\right],\\
&&\frac{dr}{d\tau}\sim\gamma\nu\sin\alpha\sin\psi,\notag\\
&&\frac{d\theta}{d\tau}\sim\frac{\gamma\nu\cos\psi}{r}.\notag
\end{eqnarray}

\section{Analysis of the critical hypersurface}
\label{sec:ACH}
We consider the implicit equation for the critical hypersuface (\ref{eq:crit_hyper2}) and using Eq. (\ref{eq:EST1}), we can determine an estimation of the critical radius $r_{\rm crit}$, i.e.,
\begin{equation}\label{eq:estcr}
\frac{M-A}{r_{\rm crit}^2}-\frac{2aM\sin\theta\nu_{\rm crit}}{r_{\rm crit}^3}-\frac{\nu^2_{\rm crit}}{r_{\rm crit}}\approx0,
\end{equation}
which turns out in solving the following quadratic algebraic equation, since $\nu_{\rm crit}=[\cos\beta]_{r=r_{\rm crit}}$, i.e.,\begin{equation}\label{eq:AE}
(M-A)\sin^2\theta r_{\rm crit}^2-b^2r_{\rm crit}-2aM\sin^2\theta b=0.
\end{equation}
The solution is given by
\begin{equation}\label{eq:SAE}
r_{\rm crit}\sim \frac{b^2}{(M-A)\sin^2\theta}\sim\frac{R^4_\star\Omega^2_\star}{(M-A)}\sin^2\theta,
\end{equation}
while the derivative $\partial_\theta r_{\rm crit}$ is given by
\begin{equation}
\partial_\theta r_{\rm crit}\sim \frac{R^4_\star\Omega^2_\star}{(M-A)}\sin(2\theta).
\end{equation}
In addition, through Eq. (\ref{eq:SAE}) we easily prove that the critical radius is finite, because
\begin{equation}\label{eq:FCR}
r_{\rm crit}\le\frac{R^4_\star\Omega^2_{\rm max}}{(M-A)}<\infty,
\end{equation}
choosing $\Omega_\star\ge\Omega_{\rm max}$, where $\Omega_{\rm max}$ is the maximum value that $\Omega_\star$ can assume (see Ref. \cite{Bakala2019}, for more details), $R_\star\le\bar{R}$, and $A/M\neq1$. Indeed, for $A/M=1$ we have that $L=L_{\rm Edd}$ and the test particle is pushed at infinity, having any critical hypersurface.

\section{Analysis of mechanical energy}
\label{sec:AKE}
To prove $\dot{\mathbb{K}}\le0$, we need to analyse the sign of each single components appearing in Eq. (\ref{eq:DLF1}). First, we note that $-(A-M)\dot{r}/r^2\le0$, because $-(A-M)/r^2\ge0$, while $\dot{r}\sim\sin\alpha\le0$. 

The term $d (\nu_{\rm crit})/d \tau$ is given by
\begin{equation}
\frac{d}{d\tau}\nu_{\rm crit}=\frac{\partial\nu_{\rm crit}}{\partial \theta}\dot{\theta}.
\end{equation}
The function $\nu_{\rm crit}$ can be so estimated (see Eq. (\ref{ANG1})) as
\begin{equation} \label{eq:BETA}
\nu_{\rm crit}\sim \frac{b}{r_{\rm crit}|\sin\theta|}.
\end{equation}
The partial derivative term $\partial\nu_{\rm crit}/\partial \theta$ is
\begin{eqnarray}
\frac{\partial\nu_{\rm crit}}{\partial \theta}&&=\left[\frac{\partial_\theta(bN)\sqrt{g_{\varphi\varphi}}(1+bN^\varphi)}{g_{\varphi\varphi}(1+bN^\varphi)^2}\right.\label{eq:PD2}\\
&&\left.-\frac{bN\partial_\theta(\sqrt{g_{\varphi\varphi}}(1+bN^\varphi))}{g_{\varphi\varphi}(1+bN^\varphi)^2}\right]_{r=r_{\rm crit}}.\notag
\end{eqnarray}
We consider the estimations of all derivatives appearing in Eq. (\ref{eq:PD2}), where we have
\begin{equation}\label{eq:DTHETA}
\begin{aligned}
&\partial_\theta \Delta\sim\partial_\theta\rho\sim2r_{\rm crit}\partial_\theta r_{\rm crit},\quad \partial_\theta N\sim-\frac{Ma^2\partial_\theta r_{\rm crit}}{r_{\rm crit}^4},\\
&\partial_\theta\sqrt{g_{\varphi\varphi}}\sim \frac{a^2\cos\theta}{r_{\rm crit}},\quad \partial_\theta N^\varphi\sim-\frac{2Ma\partial_\theta r_{\rm crit}}{r_{\rm crit}^4},\\
&b\sim R_\star^2\Omega\sin^2\theta,\quad \partial_\theta b\sim R_\star^2\Omega\sin2\theta.
\end{aligned}
\end{equation}
Therefore, we obtain
\begin{equation}
\begin{aligned}
\partial_\theta\nu_{\rm crit}&\sim\frac{1}{r_{\rm crit}\sin^2\theta}\left[r_{\rm crit}\sin\theta\partial_\theta b-\frac{ba^2\cos\theta}{r_{\rm crit}}\right]\\
&\sim \frac{\partial_\theta b}{r_{\rm crit}\sin\theta}\ge0.
\end{aligned}
\end{equation}
Since $\dot{\theta}\ge0$, we obtain $-\cos\beta d(\cos\beta)/d\tau\sim -b^2\le0$. 

We have to study only the term $\mathbb{A}=d\nu/d\tau-(A-M)\dot{r}/r^2$, which, after defined $x=\sin\psi\cos(\alpha-\beta)$,  reads explicitly as
\begin{equation}
\begin{aligned}
\mathbb{A}&=\frac{\nu A}{M}\left\{\sin\alpha\sin\psi\left(\gamma-\frac{1}{\gamma}\right)\right.\\
&\left.+\frac{A}{M}\left[\gamma\sin\alpha\sin\psi+(1-\nu x)(x-\nu)\right]\right\}.
\end{aligned}
\end{equation}
The terms $\gamma\sin\alpha\sin\psi$ and $\sin\alpha\sin\psi\left(\gamma-\frac{1}{\gamma}\right)$ are both non-positive.
We distinguish two cases:
\begin{itemize}
\item if $\nu> x$, it is obvious that $\mathbb{A}$ is non-positive, because it is the sum of three negative terms;
\item if $\nu\le x$, it is not evident, but we need to perform some calculations, where we have
\begin{equation}
\begin{aligned}
&(1-\nu x)(x-\nu)=x(1+\nu^2)-\nu(1+x^2)\\
&\le x(1+\nu^2)-\nu(1+x^2)\le0.
\end{aligned}
\end{equation}
\end{itemize}

It is important to note that the sign function ${\rm sgn}(\nu^2-\cos^2\beta)$ does not change the final sign of $\dot{\mathbb{K}}$, because the former multiplies only the kinetic term, while the negative dominant contribution from the gravitational and radiation potential persists with its own sign. Therefore, we have finally proved that $\dot{\mathbb{K}}\le0$, both on the critical hypersurface and at the equatorial ring.

\section{Analysis of angular momentum}
\label{sec:AAM}
This case relies on the results found in Sec. \ref{sec:AKE}. Considering Eq. (\ref{eq:DLF2}), we define $\dot{\mathbb{L}}\sim\mathcal{C}_2-\mathcal{C}_1$, where
\begin{eqnarray}
&&\mathcal{C}_1=\dot{r}_{\rm crit}\nu_{\rm crit}+r_{\rm crit}\frac{d(\nu_{\rm crit})}{d\tau},\\
&&\mathcal{C}_2=r\frac{d\nu}{d\tau}\cos\alpha\sin\psi+\nu(\dot{r}\cos\alpha\sin\psi\notag\\
&&\qquad-r\sin\alpha\sin\psi\ \dot{\alpha}+r\sin\alpha\cos\psi\ \dot{\psi}).
\end{eqnarray}
We start by analysing $\mathcal{C}_1$, where we obtain
\begin{equation}
\begin{aligned}
\mathcal{C}_1&\ge \partial_\theta r_{\rm crit}\dot{\theta}\left(\frac{b}{r_{\rm crit}\sin\theta}+r_{\rm crit}\right)\\
&\ge r_{\rm crit}\partial_\theta r_{\rm crit}\dot{\theta}\ge0.
\end{aligned}
\end{equation}
Therefore, we have $-\mathcal{C}_1\le0$.

Now, it is important to analyse the sign of $\mathcal{C}_2$. Employing Eqs. (\ref{eq:EST1}) -- (\ref{eq:EST2}), we have
\begin{equation}
\begin{aligned}
\mathcal{C}_2\le-\gamma\nu^2\cos^3\alpha\cot\theta\left(\sin\psi\cos\psi+\sin^2\alpha\right)\le0.
\end{aligned}
\end{equation}
Therefore, we finally have that $\dot{\mathbb{L}}\le0$, both on the critical hypersurface and at the equatorial ring.

\section{Analysis of Rayleigh potential}
\label{sec:ARP}
This case is partially based on previous calculations and new estimations. Equation (\ref{eq:DLF3}) entails to estimate the following derivatives
\begin{eqnarray}
\dot{\mathcal{I}^2}&&=\frac{\partial \mathcal{I}^2}{\partial r}\dot{r}+\frac{\partial \mathcal{I}^2}{\partial \theta}\dot{\theta},\label{eq:D31}\\
\dot{\mathbb{E}}&&=\frac{\partial \mathbb{E}}{\partial\nu}\dot{\nu}+\frac{\partial \mathbb{E}}{\partial\psi}\dot{\psi}+\frac{\partial \mathbb{E}}{\partial\alpha}\dot{\alpha}+\frac{\partial \mathbb{E}}{\partial r}\dot{r}+\frac{\partial \mathbb{E}}{\partial\theta}\dot{\theta},\label{eq:D32}\\
\dot{\mathbb{E}}_{\rm crit}&&=\frac{\partial \mathbb{E}_{\rm crit}}{\partial\theta}\dot{\theta}.\label{eq:D33}
\end{eqnarray}

First, we consider Eq. (\ref{eq:D31}), where we perform the following estimations
\begin{eqnarray}
\frac{\partial \mathcal{I}^2}{\partial r}&&\sim -\frac{\mathcal{I}^2_0\left(2r^2-\frac{b^2}{\sin^2\theta}\right)}{r^2\left(r^2-\frac{b^2}{\sin^2\theta}\right)^{3/2}},\\
\frac{\partial \mathcal{I}^2}{\partial \theta}&&\sim\frac{\mathcal{I}^2_0}{r\left(r^2-\frac{b^2}{\sin^2\theta}\right)^{3/2}}\frac{R_\star^4\Omega_\star^2\sin(2\theta)}{2}.
\end{eqnarray}
It is important to note that since the square root function appearing at the denominator of Eq. (\ref{INT_PAR}) it is well defined, when its argument is positive, therefore we conclude that $r^2-b^2/\sin^2\theta>0$. Therefore, we have
\begin{equation}
\begin{aligned}
\dot{\mathcal{I}^2}&\sim\frac{\mathcal{I}^2_0}{r\left(r^2-\frac{b^2}{\sin^2\theta}\right)^{3/2}}\frac{R_\star^4\Omega_\star^2\sin(2\theta)}{2}\ge0.
\end{aligned}
\end{equation}

Then, we pass to analyse Eq. (\ref{eq:D32}), estimating the following partial derivatives, i.e.,
\begin{eqnarray}
\frac{\partial \mathbb{E}}{\partial\nu}&&\sim\gamma^3 E_p[\nu-\sin\psi\cos(\alpha-\beta)],\\
\frac{\partial \mathbb{E}}{\partial\psi}&&\sim -\gamma E_p\nu\cos\psi\cos(\alpha-\beta),\\
\frac{\partial \mathbb{E}}{\partial\alpha}&&\sim\gamma E_p\nu\sin\psi\sin(\alpha-\beta),\\
\frac{\partial \mathbb{E}}{\partial r}&&\sim-\frac{M\gamma E_p}{r^2},\\
\frac{\partial \mathbb{E}}{\partial\theta}&&\sim\frac{Ma\gamma E_p\sin(2\theta)}{r^3}(a-2R^2_\star\Omega).
\end{eqnarray}
Combinanig such results, together with Eqs. (\ref{eq:EST1})--(\ref{eq:EST2}), we obtain
\begin{equation}
\begin{aligned}
\dot{\mathbb{E}}&\sim\frac{\gamma^2 E_p\nu^2}{r}\left\{
\cos^2\psi\sin\alpha\left[\cos(\alpha-\beta)-\sin\alpha\cos\alpha\cot\theta\right]\right.\\
&\left.-\cos\psi\sin\psi\cos(\alpha-\beta)\cos^2\alpha\cot\theta\right.\\
&\left.-\sin(\alpha-\beta)\cos\alpha\right\}\ge0.
\end{aligned}
\end{equation}
Since $\mathbb{E}\ge0$ (because it is an energy, always defined non-negative), we infer that $-\dot{\mathbb{E}}/\mathbb{E}\le0$.

Finally, it remains only to estimate Eq. (\ref{eq:D33}). It is useful to perform the following estimation
\begin{eqnarray}
\partial_\theta \sin\beta\sim-\cot\beta\partial_\theta\cos\beta.
\end{eqnarray} 
We need to calculate and estimate the partial derivative $\partial_\theta\mathbb{E}_{\rm crit}$. Using such estimation together with Eqs. (\ref{eq:EST1})--(\ref{eq:EST2}) and Eqs. (\ref{eq:DTHETA}), we obtain
\begin{equation}
\begin{aligned}
\partial_\theta\mathbb{E}_{\rm crit}&\sim-E_p\frac{(\cos\beta)_{\rm crit}}{|(\sin\beta)_{\rm crit}|}\frac{\partial_\theta b}{r_{\rm crit} \sin\theta}
\le0.
\end{aligned}
\end{equation}
Since $\dot{\theta}\ge0$ and $\mathbb{E}_{\rm crit}\ge0$, we have $\dot{\mathbb{E}}_{\rm crit}/\mathbb{E}_{\rm crit}\le0$.

We note that $\dot{\mathbb{E}}/\mathbb{E}\sim1/r$, dominates over $\dot{\mathcal{I}^2}\sim1/r^3$. Therefore,   
we conclude that $\dot{\mathbb{F}}\le0$, both on the critical hypersurface and at the equatorial ring.

\end{appendix}

\bibliography{references}

\end{document}